# Electronic Structure and Stability of the $CH_3NH_3PbBr_3$ (001) Surface

*Xin Huang[1,2], Tula R. Paudel[1,\*], Peter A. Dowben[1], Shuai Dong[2], and Evgeny Y. Tsymbal[1,\*]*

[1]Department of Physics and Astronomy & Nebraska Center for Materials and Nanoscience,

University of Nebraska, Lincoln, Nebraska 68588, USA

[2]Department of Physics, Southeast University, Nanjing 211189, China

The energetics and the electronic structure of methylammonium lead bromine ($CH_3NH_3PbBr_3$) perovskite (001) surfaces are studied based on density functional theory. By examining the surface grand potential, we predict that the $CH_3NH_3Br$-terminated (001) surface is energetically more favorable than the $PbBr_2$-terminated (001) surface, under thermodynamic equilibrium conditions of bulk $CH_3NH_3PbBr_3$. The electronic structure of each of these two different surface terminations retains some of the characteristics of the bulk, while new surface states are found near band edges which may affect the photovoltaic performance in the solar cells based on $CH_3NH_3PbBr_3$. The calculated electron affinity of $CH_3NH_3PbBr_3$ reveals a sizable difference for the two surface terminations, indicating a possibility of tuning the band offset between the halide perovskite and adjacent electrode with proper interface engineering.

*tula.paudel@gmail.com, tsymbal@unl.edu

# I.   Introduction

The remarkable performance and rapid growth of the power conversion efficiencies of the Organic-inorganic hybrid perovskites has been stunning. [1,2,3,4,5] First attempts of using halide perovskites $CH_3NH_3PbX_3$ (X = I, Br) as a light absorber in liquid-based dye-sensitized solar cell demonstrated the efficiency of 3.8% for X = I and 3.1% for X = Br.[6] Within 5-6 years, the power conversion efficiency of these extraordinary materials has soared to reach 20%.[5,7,8,9,10] Very recently, it was discovered that the perovskite solar cells have facet-dependent photovoltaic efficiencies, with a highest photocurrent reaching the theoretical conversion limit.[11] This discovery opens the door to engineer the material's facets to further boost the photovoltaic efficiency of solar cells.

These remarkable developments are due to unique photovoltaic properties of the organo-lead halide perovskites, superior to those exploited in the conventional dye-sensitized solar cells. A primary advantage of these hybrid organo-lead materials is the significantly greater optical absorption compared to the conventional thin-film solar cell absorbers, enabling the use of ultrathin films in collecting photogenerated carriers, with the benefit of a lower non-radiative recombination rate.[2,12,13,14,15,16,17] In addition, the organo-lead halide perovskites show long electron-hole diffusion lengths (exceeding 1 micrometer) and high mobilities for both electron and hole charge carriers, resulting in ambipolar charge transport.[18,19,20,21,22,23] The band gap (in the region of 2 eV) and shallow defect levels, even at the surface, also contribute to their remarkable photovoltaic performance.[24,25,26]

Typically, a perovskite-based solar cell represents a layered structure with $TiO_2$ serving as an electron transport layer, the $CH_3NH_3PbX_3$ layer acting as an optical harvester, and a hole transport layer. Not surprisingly, the characteristics of surfaces and interfaces are expected to play a key role in the performance of the perovskite solar cell. While there have been a number theoretical efforts devoted to studies of bulk $CH_3NH_3PbX_3$ materials, there are only a few theoretical reports discussing properties of $CH_3NH_3PbX_3$ surfaces or interfaces. The latter mostly focus on the structural stability of $CH_3NH_3PbI_3$ surfaces of different orientations [27,28,29,30] or properties of the $CH_3NH_3PbI_3$/$TiO_2$ interfaces.[31,32,33]

In this paper, we explore the surface properties of the $CH_3NH_3PbBr_3$ perovskite photovoltaic material. Compared to $CH_3NH_3PbI_3$, $CH_3NH_3PbBr_3$ has a larger bandgap of about 2.2 eV and thus a lower optical absorption, but produces a higher open-circle voltage. [34,35,36,37] An open circle voltage as high as 1.5 eV with 10.4% power conversion efficiency has been realized in a $CH_3NH_3PbBr_3$-based solar cell.[35] A similar open circuit voltage has also been obtained in a $CH_3NH_3PbBr_{3-x}Cl_x$-based solar cell.[36] Furthermore, mixing Br into $CH_3NH_3PbI_3$ reduces the charge recombination rates in the light absorber film, which improves the solar cell stability and enhances the lifetime of the device.[38,39] The variable bandgap of this mixed halide perovskite can also be used for application in a multi-junction solar cell. Although there remains much room for improvement in the available $CH_3NH_3PbBr_3$-based solar cells, compared to the $CH_3NH_3PbI_3$-based solar cells, the $CH_3NH_3PbBr_3$ perovskite is considered to be a suitable candidate for obtaining superior photovoltaic performance.

We focus on the electronic properties, especially surface states, of different terminations of the cubic $CH_3NH_3PbBr_3(001)$ surfaces. From the viewpoint of growth conditions, the surface termination is reflected in different chemical potentials, which is accessible from density functional theory (DFT). By comparing the surface grand potential, we predict the favorable surface termination and explore the surface states. Finally, we calculate the electron affinity of $CH_3NH_3PbBr_3$ for different surface terminations and discuss implications of our results from the viewpoint of photovoltaic efficiency.

## II.  Computation Methods and Atomic Structure

Our computational approach employs a plane-wave pseudopotential method, as implemented in the Vienna *ab initio* Simulation (VASP).[40,41] The exchange-correlation functional is treated within the Perdew-Burke-Ernzerhof (PBE) generalized gradient approximation (GGA).[42,43] For the simulation of surface, a plane-wave cutoff energy of 660 eV and a $8 \times 8 \times 1$ Monkhorst-Pack k-grid mesh is used for the total-energy and atomic relaxation calculations. The in-plane lattice constants are constrained to the optimized bulk cubic $CH_3NH_3PbBr_3$ parameters, while the whole atomic positions are relaxed until the Hellmann-Feynman forces on each atom become less than 20 meV/Å. We ensure the convergence of the results with respect to the supercell size by make sure that the surface induced relaxation dies down at the middle of the supercell, and results from a slightly smaller supercell are qualitatively similar. The surface induced relaxation increases the c-lattice constant to 6.15 Å for the $CH_3NH_3Br$ surface termination and deceases to 5.95 Å for the $PbBr_2$ surface termination. This surface layer relaxation do, however, vanishes beyond ~2 nm away from the surface so that the middle cell of the slab retains bulk-like c-lattice constant of ~ 6.08 Å. In addition, we obtained qualitatively similar results for the smaller 9-layered $PbBr_2$ terminated slab and the 11-layered $CH_3NH_3Br$ slab. All the calculations have taken into account the spin-orbit interaction. We include a dipole layer in the vacuum, to eliminate unphysical electric field in the direction perpendicular to the slab.

At room temperature, bulk $CH_3NH_3PbBr_3$ exhibits a cubic structure with the $Pm\bar{3}m$ space group symmetry. Our optimized lattice parameter $a$ equals to 6.08 Å, which is a slight overestimation with respect to the experimental value of 5.94 Å.[39] In $CH_3NH_3PbBr_3$, the $Pb^{2+}$ ions are surrounded by six $Br^-$ ions forming $PbBr_6$ octahedrons and the dipolar organic methylammonium ($CH_3NH_3^+$) ions are located in the octahedron $PbBr_6$ cages (see Fig. 1(a)). An asymmetric charge distribution within the $CH_3NH_3$ cation, with the positive charge localized mainly on the $NH_3$ group, results in an uncompensated electric dipole moment oriented from the carbon to the nitrogen atom. Experimentally, the dipole moments of $CH_3NH_3$ are randomly oriented causing $CH_3NH_3PbBr_3$ to be paraelectric at room temperature.[44,45] In order to simulate the paraelectric character of bulk $CH_3NH_3PbBr_3$, we assume that the dipole moments of the methyl ammonium in two consecutive (i.e. adjacent) atomic planes along the [001] axis are pointing in the opposite directions along the [111] axis. This doubles the cubic unit cell in our calculations, as depicted in Fig. 1(a). The relaxed atomic positions in such a non-polar (paraelectric) cell are closer to ideal $Pm\bar{3}m$ than in the polar structure. This constraint on the lattice and the methyl ammonium orientation ensures that the net dipole moment in the whole unit cell is fully compensated due to the opposite direction of the two $CH_3NH_3$ dipoles.

To explore the influence of the surface, we construct a symmetric supercell stacked along the [001] direction. Along this direction, the $CH_3NH_3PbBr_3$ crystal consists of alternating neutral $CH_3NH_3Br$ and $PbBr_2$ planes, allowing for two possible surface terminations of the slab, i.e. either the $CH_3NH_3Br$ or the $PbBr_2$ surface termination. In order to model the (001) surface, with the $CH_3NH_3Br$ or $PbBr_2$ surface terminations, we construct $CH_3NH_3PbBr_3$ slabs consisting of $1\times1\times7$ or $1\times1\times6$ cubic unit cells, respectively, as shown in Fig. 2. These slabs are separated by a vacuum region of 10 Å to ensure that the surfaces of the periodically repeated slabs are well decoupled. For each surface termination, we consider two different structures distinguished by an opposite dipole orientation of the $CH_3NH_3$ cations (compare Figs. 2(a) and 2(c) to Figs. 2(b) and 2(d), respectively). We denote these different structures as "dipole-into" and "dipole-away" when the methylammonium dipole is pointing into the first $PbBr_2$ layer (Fig. 2(a) and 2(c)) and when the dipole is pointing away from the first $PbBr_2$ layer (Fig. 2(b) and 2(d)) respectively.

## III. Results and Discussion

### A. Surface Termination

Since, there are two possible dipole orientations for each termination (corresponding to the "dipole-into" and the "dipole-away" slabs), we have first determined the ground state for both the $CH_3NH_3Br$- and $PbBr_2$-terminated surfaces. The results indicate that the total energy of the "dipole-into" slabs is lower than that of the "dipole-away" slabs for both surface terminations. Specifically, the calculated energy difference is 3.34 meV/atom for the $PbBr_2$-terminated surface and 8.61 meV/atom for the $CH_3NH_3Br$-terminated surface. Hence, below we focus on the structural stability of the "dipole-into" slabs.

To compare the relatively structural stability of the two surface terminations, we compute the surface grand potential for each termination. We introduce the chemical potentials $\mu_{CH_3NH_3}$, $\mu_{Pb}$, and $\mu_{Br}$ for the CH₃NH₃ cation, Pb and Br atomic species, respectively, and write down the surface grand potential per unit area, $\Omega^i$, for the $i$ termination as follows:

$$\Omega^i \approx \frac{1}{2S}[E_{CH_3NH_3PbBr_3}^{slab} - N_{CH_3NH_3}\mu_{CH_3NH_3} - N_{Pb}\mu_{Pb} - N_{Br}\mu_{Br}]. \qquad (1)$$

Here $E_{CH_3NH_3PbBr_3}^{slab}$ refers to the total energy of the slab supercell, $N_{CH_3NH_3}$, $N_{Pb}$ and $N_{Br}$ are the number of CH₃NH₃ cations, Pb and Br atoms in the slab, respectively. The factor of 1/2 takes into account the existence of the two identical surfaces in the slab where $S$ is the surface area. The chemical potential $\mu_{CH_3NH_3PbBr_3}$ of a stoichiometric CH₃NH₃PbBr₃ phase is given by sum of the three terms, representing the chemical potentials of each atomic constituent, within the crystal, as follows:

$$\mu_{CH_3NH_3PbBr_3} = \mu_{CH_3NH_3} + \mu_{Pb} + 3\mu_{Br}. \qquad (2)$$

Since the surface is in equilibrium with the bulk CH₃NH₃PbBr₃, we have $\mu_{CH_3NH_3PbBr_3} = E_{CH_3NH_3PbBr_3}^{bulk}$. Using this relationship and Eq. (2), we may eliminate the $\mu_{CH_3NH_3}$ variable in Eq. (1) so that it becomes:

$$\Omega^i \approx \frac{1}{2S}\Big[E_{CH_3NH_3PbBr_3}^{slab} - N_{CH_3NH_3}E_{CH_3NH_3PbBr_3}^{bulk} - (N_{Pb} - N_{CH_3NH_3})\mu_{Pb} - (N_{Br} - 3N_{CH_3NH_3})\mu_{Br}\Big], \quad (3)$$

Relying upon Eq. (3), one can then deduce the range of the accessible values of the surface grand potential per unit area $\Omega^i$ for each termination, if the minimum and maximum values of the Pb and Br chemical potentials are known. In our calculations, the allowable chemical potentials for Pb and Br should satisfy the existence of bulk CH₃NH₃PbBr₃ in thermodynamic equilibrium growth conditions (as reflected by Eq. (4) below), and also avoid the formation of possible secondary phases of $CH_3NH_3Br$ and $PbBr_2$ (as reflected by Eqs. (5) and (6) below). These conditions are given by the following relationships:[46]

$$\Delta\mu_{CH_3NH_3} + \Delta\mu_{Pb} + 3\Delta\mu_{Br} = \Delta H_{\text{form}}(CH_3NH_3PbBr_3) = -6.780\text{eV}\,, \tag{4}$$

$$\Delta\mu_{CH_3NH_3} + \Delta\mu_{Br} < \Delta H_{\text{form}}(CH_3NH_3Br_2) = -3.633\text{eV}\,, \tag{5}$$

$$\Delta\mu_{Pb} + 2\Delta\mu_{Br} < \Delta H_{\text{form}}(PbBr_2) = -3.061\text{eV}\,. \tag{6}$$

Here, $\Delta\mu$ represents the variation of the chemical potential, with respect to those computed for the reference phases, e.g. $\Delta\mu_{Pb} = \mu_{Pb} - \mu_{Pb}^{Bulk} = \mu_{Pb} - E_{Pb}^{Bulk}$, $\Delta\mu_{Br} = \mu_{Br} - \mu_{Br}^{mol} = \mu_{Br} - E_{Br_2}^{mol}/2$, indicating the environmental conditions. In Eqs. (4), (5) and (6), $\Delta H_{\text{form}}$ is the heat of formation, which is calculated as the difference between the total energy of the compound and the composition-weighted sum of their constituents, i.e. $\Delta H_{\text{form}}[AB] = E_{tot}[AB] - E_{tot}[A] - E_{tot}[B]$. Values of the total energy and heat of formation, $E_{\text{tot}}$ and $\Delta H_{\text{form}}$, of solid and gas phases mentioned above, have been listed in Table 1.

The range of the chemical potentials satisfying Eqs. (4), (5) and (6) is depicted by the red region in Fig. 3. The right upper and left lower edges of this range represents the borders to the $PbBr_2$- and $CH_3NH_3Br$-rich conditions, respectively. From Fig. 3, it is evident that the possible chemical potential range for stable growth of bulk $CH_3NH_3PbBr_3$ is very narrow, consistent with the recent results by Shi et al.[24] This narrow thermodynamic stable range for equilibrium growth indicates that the cubic $CH_3NH_3PbBr_3$ compound easily decomposes into $CH_3NH_3Br$ and $PbBr_2$, and agrees with the small dissociation energy (of only 0.19 eV) reported for $CH_3NH_3PbBr_3$ decomposition to $CH_3NH_3Br$ and $PbBr_2$.[24]

Using the accessible values for the Pb and Br chemical potentials, obtained from Eqs. (4)-(6), we can calculate the surface grand potentials for each termination (i.e. $CH_3NH_3Br$ or $PbBr_2$) using Eq. (3). The results are displayed in Fig. 3, showing the regions where the $CH_3NH_3Br$ and $PbBr_2$ surface termination are stable. It is evident that the whole chemical potential range, where growth of the equilibrium bulk $CH_3NH_3PbBr_3$ is possible, is located in the region where the $CH_3NH_3Br$ surface termination is thermodynamically more stable than the $PbBr_2$ termination for the $CH_3NH_3PbBr_3$ (001) surface. The conclusion is that the $CH_3NH_3Br$ surface termination is favored. This predicted surface termination is confirmed by the experiment [47] and resembled that found for $CH_3NH_3PbI_3$.[48]

To compare the surface energy differences between the $CH_3NH_3Br$ or $PbBr_2$ surface terminations, we choose three representative points from the accessible chemical potential range in Fig. 3. The points are labeled as A (Pb-rich/Br-poor, $\mu_{CH_3NH_3} = -2.058$ eV, $\mu_{Pb} = 0$, $\mu_{Br} = -1.574$ eV), B (Pb-moderate/Br-moderate, $\mu_{CH_3NH_3} = -2.879$ eV, $\mu_{Pb} = -1.522$ eV, $\mu_{Br} = -0.793$ eV) and C (Pb-poor/Br-rich, $\mu_{CH_3NH_3} = -3.633$ eV, $\mu_{Pb} = -3.147$ eV, $\mu_{Br} = 0$). As seen from Table 2, the surface energy differences between $CH_3NH_3Br$ and $PbBr_2$ terminations are small and lie within 5 meV/Å$^2$. Therefore, we conclude that the $CH_3NH_3Br$ and $PbBr_2$ terminations may coexist on a single surface, with the latter being dominant under the $PbBr_2$-rich condition (the pink area in Fig. 3). In fact, the surface energy differences at the termination border ($\mu_{CH_3NH_3} = -2.55$ eV, $\mu_{Pb} = 0$, $\mu_{Br} = -1.41$ eV and $\mu_{CH_3NH_3} = -3.98$ eV, $\mu_{Pb} = -2.8$ eV, $\mu_{Br} = 0$) are also small, at a value of about 4 meV/Å$^2$.

## B. The Electronic Band Structure

Fig. 4 shows the calculated electronic band structures of the $CH_3NH_3PbBr_3$(001) surface, for the different $CH_3NH_3Br$ (Fig. 4(a)) and $PbBr_2$ (Fig. 4(b)) surface terminations. Both surfaces show a direct band gap located at the M point of the Brillouin zone. No midgap states are found for either termination, indicating no intrinsic mechanism to support electron-hole recombination at the surfaces. Suppression of electron-hole recombination is beneficial for realizing a large diffusion length of photoexcited carriers in $CH_3NH_3PbBr_3$ perovskites. Similar to the bulk phase, the surface conduction and valence bands consist mostly of Pb 6p and Br 5p orbitals, respectively. A high carrier mobility is expected in the $CH_3NH_3PbBr_3$ surfaces due to their dispersive bands occurring at the conduction band minimum (CBM) and valence band maximum (VBM). Our calculated effective masses for electrons and holes at the CBM and VBM are $m_e = 0.23\ m_0$ and $m_h = 0.24\ m_0$ for $CH_3NH_3Br$ termination and $m_e = 0.23\ m_0$ and $m_h* = 0.26\ m_0$ for $PbBr_2$ termination, respectively. These values are comparable to those found in silicon ($m_e* = 0.19\ m_0$ and $m_h* = 0.16\ m_0$) and consistent with the ambipolar carrier transport behavior revealed in perovskite solar cells.

The band gap is a crucial material parameter for photovoltaic applications. Our calculations cannot predict quantitatively the magnitude of the band gap, due to the well-known deficiency of the GGA method. [31,38,17,20] However, we can make conclusions about changes in the band gap depending on the interface termination. Table III shows that the calculated values significantly underestimate the experimental band of about 2.3 eV.[20] It is notable that the magnitude of the bulk band gap depends on whether the organic perovskite is considered to be polar or non-polar. For a uniformly polarized bulk $CH_3NH_3PbBr_3$ crystal, the band gap appears to be about 0.27 eV larger than that for a non-polar $CH_3NH_3PbBr_3$ crystal. This result can be understood from the layer-resolved local density of states (LDOS) calculated for non-polar bulk $CH_3NH_3PbBr_3$, as shown in Fig. 1(b). The CBM is located at the middle $PbBr_2$ monolayer, where the $CH_3NH_3^+$ dipoles from the adjacent layers meet with a head-to-head configuration, resulting in accumulated positive bound charge. Contrary, the VBM is located at the bottom $PbBr_2$ layer with a tail-to-tail configuration of the $CH_3NH_3$ dipoles with accumulated negative bound charges. This bound charge variation along the z-direction gives rise to an alternating electrostatic potential, leading to the downshift of the CBM and upshift of the VBM and thus leading to a reduced band gap. Additionally, we find large octahedral distortions in polar bulk crystal compared to the non-polar bulk $CH_3NH_3PbBr_3$ crystal. The measured Pb-Br-Pb bond angles in $PbBr_3$ octahedra is 175 ° in polar bulk $CH_3NH_3PbBr_3$ compared to 179 ° in non-polar bulk $CH_3NH_3PbBr_3$ crystal. The large octahedral distortion in polar structure reduces overlap between the Br $p$-orbitals and the Pb $s$-orbitals, and might also be responsible for the wider band gap in the polar bulk crystal. Indeed, the band gap can be controlled by varying the degree of octahedral distortion through external means such as epitaxial strain, as Grote et al. found that increasing the octahedral distortion can widen the band gap in a tin-halide perovskite.[49] The same mechanisms are valid in our slab calculations, where the electrostatic potential periodically changes along the z-direction resulting in the band gap similar to that for the non-polar bulk $CH_3NH_3PbBr_3$. We note that in practice, where we expect a random orientation of the $CH_3NH_3^+$ dipoles, the electrostatic potential is thus randomly varied from cell to cell resulting in the similar mechanism of the band gap reduction with respect to a uniformly polarized system.

We also explore the relative weight of the surface bands for each termination. Here, we treat the contribution from one top and one bottom unit cells in the slab as having surface character (colored in blue) and the contribution from the rest part of the slab as retaining the bulk electronic structure (colored in red). As seen from Fig. 4, the surface states show distinct features for both terminations. For the $CH_3NH_3Br$ termination, the CBM has a large surface contribution and there are more surface weighted states located deeper in the valence band ($< -1$ eV). We find that the states near the CBM are strongly affected by the spin-orbit interaction resulting in level splitting and hybridization between the surface and bulk weighted bands. This is different from the $PbBr_2$ termination, where no surface bands are found near the CBM; they are located at higher energies, well into the conduction band ($> 1.2$ eV above the CBM) and lower energies ($< -0.5$ eV) in the valence bands. The bands near the VBM are mostly of bulk character for both surface terminations. These bands and their dispersion away from the $\overline{M}$ point are consistent with the photoemission measurements.[47] We argue that the surface weighted states at the CBM in the $CH_3NH_3Br$-terminated structure may be advantageous for extraction of photoexcited electrons from the photoactive perovskite layer to the adjacent electrode. As the surface weighted states do not fall into a gap of the projected bulk band structure, they are not true surface states, but rather might be characterized as surface resonances.

## C. Ionization Potential

The efficiency of charge transfer in photovoltaic devices is also determined by the band alignment between the photoactive layer and the electrodes. The ionization potential and the electron affinity of the perovskite are important characteristics which have a significant influence on the band alignment. We calculate the ionization potential (ionization energy) of the $CH_3NH_3PbBr_3(001)$ surfaces as the difference between the vacuum energy and the VBM energy. The former is obtained from the electrostatic potential in our supercell calculation (as shown in Fig. 5). We find that there is a sizable difference between the ionization potential for the two surface terminations. The calculated ionization potential is 4.67 eV for the $CH_3NH_3Br$-terminated surface and 5.53 eV for the $PbBr_2$-terminated surface. For comparison, work function analysis of spin-coated polycrystalline films $CH_3NH_3PbBr_3$, by photoelectron spectroscopy, shows the valence-band offset is about 5.38 eV with respect to the vacuum level.[6] This experimental value lies somewhat in between of the calculated values, but does not take into account the tendency for Pb surface segregation seen elsewhere in experiment.[47] Furthermore, the ionization potentials obtained from generalized gradient approximation cannot be directly compared to the experimental ionization potentials, due to the wrong asymptotic behavior of the corresponding Kohn-Sham potentials however, due to relative error cancelations, the trends obtained in such calculation may be compared with experiment.

Our calculations provide an important message: the different surface/interface terminations produce very different ionization potentials. This is also known for other organic systems.[50] In our case, comparing minimum energy "dipole-into" structures in Fig. 1(a) and Fig. 1(c), we see that the $CH_3NH_3Br$ terminated slab is different from the $PbBr_2$ terminated slab by the presence of an additional $CH_3NH_3Br$ dipole layer on the surface of the latter. This dipole layer is responsible for a calculated ionization potential difference of 0.86 eV between the $CH_3NH_3Br$ and $PbBr_2$ surface terminations. Interestingly, we find that the sign of the ionization energy change is opposite to what is expected for the change in the work function of a metal when a dipole layer is deposited on the surface of that metal. For the dipole moment pointing into the bulk, an increase of the work function of the metal is anticipated due to potential step at the interface shifting the Fermi energy down with respect to the vacuum potential. In our case, however, the situation is opposite: the dipole layer pointing into the slab (Fig. 2(a)) reduces the difference in the electrostatic energy between the VBM and vacuum.

To explain this behavior, we have compared the electrostatic potential energy distribution across the $CH_3NH_3PbBr_3$ slab for both the $CH_3NH_3Br$ terminated (Fig. 5(a)) and $PbBr_2$ terminated (Fig. 5(b)) surfaces. It is notable that in both cases the potential energy minima corresponding to the $PbBr_2$ monolayers vary in oscillatory fashion across the slab. This behavior reflects the alternating direction of the $CH_3NH_3^+$ dipoles between the monolayers, thus resulting in periodic steps of the potential up and down. The surface $PbBr_2$ monolayer, for the $PbBr_2$ terminated slab (Fig. 5(b)), reveals an electrostatic potential energy of –8.64 eV (in its minimum), which is placed at about 14.17 eV below the vacuum energy (5.53 eV). When the dipole layer is placed on the surface, for the $CH_3NH_3Br$ terminated slab (Fig. 5(a)), the electrostatic potential energy of the subsurface $PbBr_2$ monolayer shifts down to –9.47 eV. Simultaneously the vacuum energy shifts down to 4.67 eV so that the energy difference between the subsurface $PbBr_2$ electrostatic potential and vacuum energies remain nearly unchanged (14.17 eV). This behavior implies that placing the $CH_3NH_3Br$ monolayer on the surface effectively does not produce an additional potential step but rather shifts the vacuum level down in energy with respect to the core levels, by the value equal to the change in the ionization potential, i.e. 0.86 eV. Note that the potential minima for the interior $PbBr_2$ monolayers remain nearly unchanged with placing the dipole layer of the surface.

We can estimate the magnitude of the potential step expected from placing a $CH_3NH_3^+$ molecular layer on the surface. Using the dipole moment of the $CH_3NH_3$ molecule of 2.29 Debye,[51] the surface dipole density, and taking into account the dipole orientation along the [111] direction, we can ascertain the $z$ component of the surface dipole density $D_z$. The potential step is given by $\Delta E_D = \dfrac{\varepsilon + \varepsilon_0}{2\varepsilon\varepsilon_0} e D_z$, where $\varepsilon_0$ and $\varepsilon$ are the electric permittivities of vacuum and bulk

$CH_3NH_3PbBr_3$, respectively. Using the calculated value of $\varepsilon = 15.5\varepsilon_0$, we find $\Delta E_D \approx 0.72$ eV, which is nearly identical to the downshift of the potential energy of the subsurface $PbBr_2$ layer, or shift in vacuum energy, after placing a $CH_3NH_3Br$ surface layer. We argue that putting the molecular dipole layer on the surface results in the electronic charge redistribution due to the bonding effects, which enhances the potential drop between the surface and subsurface $PbBr_2$ monolayers (compare the respective dips in the potential in Figs. 5(a) and 5(b)). Thus, given the fact that the vacuum level preserves its energy position with respect to the surface $PbBr_2$ monolayer and the core levels do not change much, the downshift in the potential energy of the surface $PbBr_2$ after placing the $CH_3NH_3Br$ layer implies the reduction of the ionization potential.

We reiterate that significant variation in the ionization potential for different surface terminations (and likely surface orientations) has important implications for photovoltaic devices. As the ionization potential determines the barrier for charge-carrier injection into the absorber layer, the energy difference between the Fermi level of adjacent metallic electrode and the conducting states of the absorber layer is an important factor controlling the photovoltaic device performance. Interface engineering may be useful to tune the band offset between the halide perovskite and adjacent electrode to enhance the photovoltaic efficiency.

## IV. Summary

In summary, the energetics and electronic properties of various $CH_3NH_3PbBr_3(001)$ surfaces have been studied using first-principles density functional theory calculations. We find that the $CH_3NH_3Br$-terminated surface is thermodynamically more stable than the $PbBr_2$-terminated surface under the thermodynamic equilibrium conditions of bulk $CH_3NH_3PbBr_3$. Surface states are found near band edges for the $CH_3NH_3Br$-terminated surface which may be beneficial for extraction of photoexcited electrons from the photoactive perovskite layer to the adjacent electrode. The calculated ionization potential of $CH_3NH_3PbBr_3$ shows a sizable difference for the two surface terminations, indicating a possibility of tuning the band offset between the halide perovskite and adjacent electrode with proper interface engineering. These results may be useful to further improve the performance of photovoltaic devices based on organic halide perovskites.

## Acknowledgements

This work was supported by the National Science Foundation through the Nebraska Materials Science and Engineering Center (MRSEC, Grant No. DMR-1420645) and the Nebraska Center for Energy Science Research. X. H. acknowledges financial support from the China Scholarship Council.

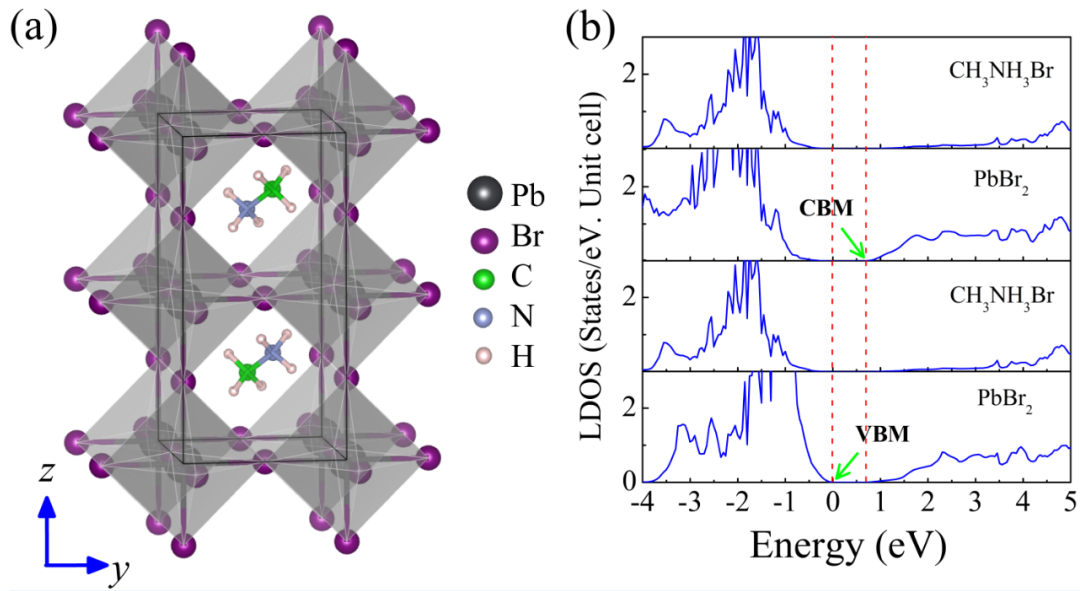

Fig. 1. (a) The optimized cubic CH₃NH₃PbBr₃ perovskite structure. The CH₃NH₃ dipole is assumed to be pointing in opposite directions in adjacent atomic layers along the [001] direction to ensure the whole structure is nonpolar. (b) The layer-resolved density of state (LDOS) as a function of energy along the $z$ direction in nonpolar bulk CH₃NH₃PbBr₃.

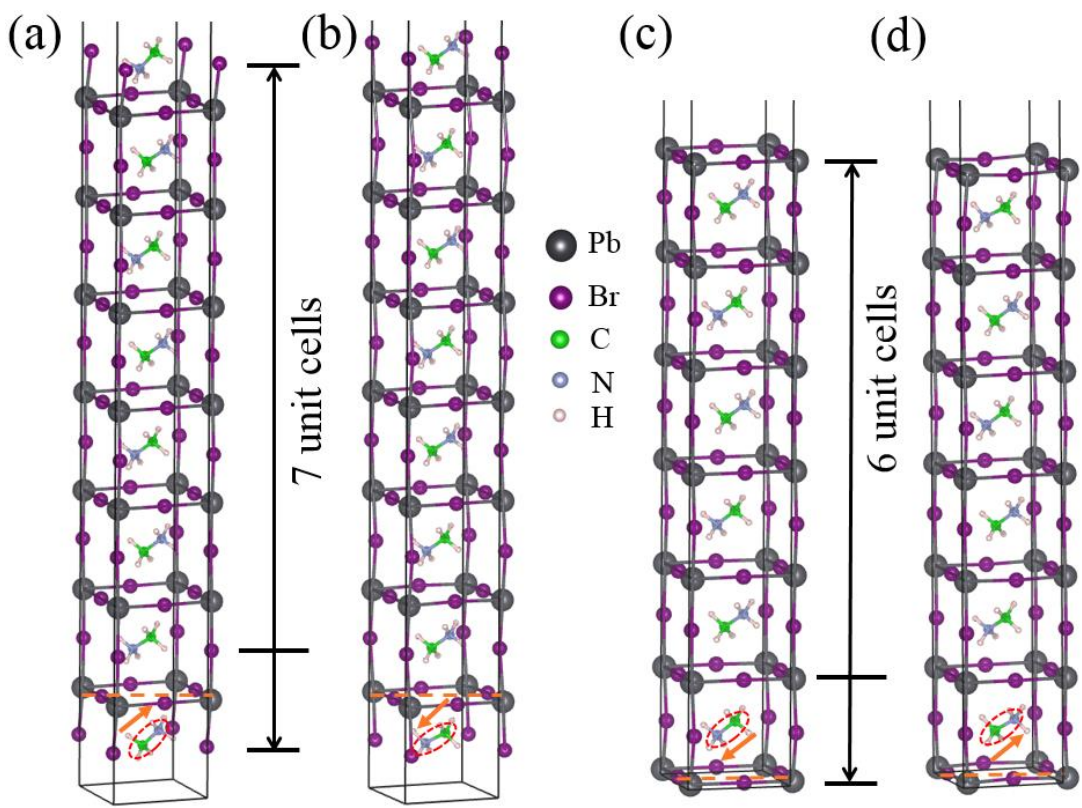

Fig. 2. The atomic structure of CH₃NH₃PbBr₃(001) slabs with CH₃NH₃Br (a, b) and PbBr₂ (c, d) surface terminations. There are two surfaces for each termination, to account for the opposite orientations of the CH₃NH₃ moieties (dipoles): (a, c) has the dipole is pointing into the first PbBr₂ layer (the "dipole-into" slab); (b, d) has the dipole is pointing away from the first PbBr₂ layer (the "dipole-away" slab). The first PbBr₂ layer is marked using orange horizontal dotted line and the dipole orientations are indicated by orange arrows.

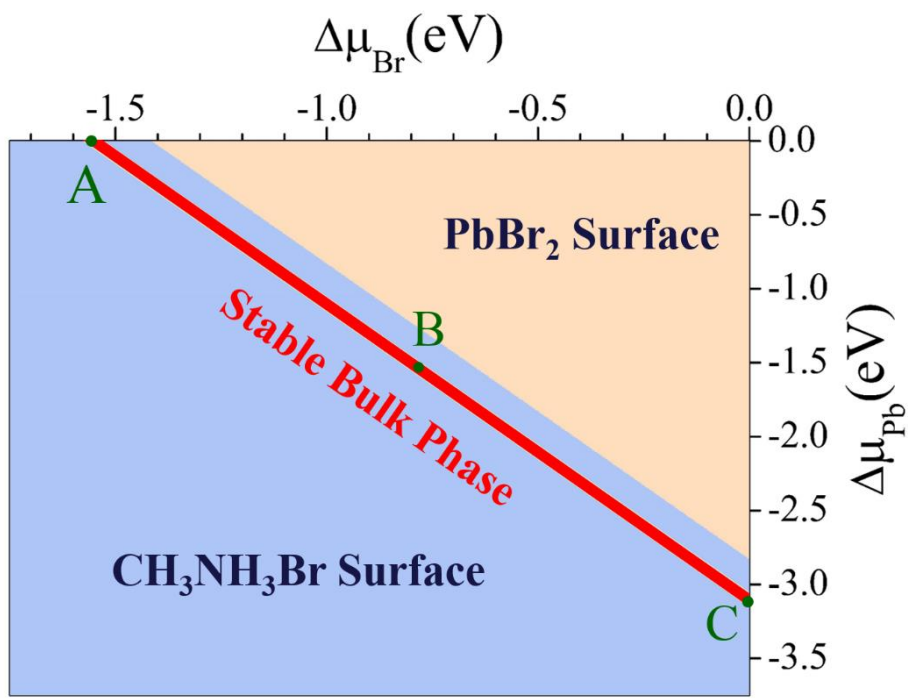

Fig. 3. The surface stability diagram for $CH_3NH_3PbBr_3$. The blue (pink) region indicates where the $CH_3NH_3Br$ ($PbBr_2$) surface termination is stable. The narrow red region is the stable range of chemical potentials for growth of bulk $CH_3NH_3PbBr_3$ in thermodynamic equilibrium. Three representative points A (Pb-rich/Br-poor), B (Pb-moderate/Br-moderate) and C (Pb-poor/Br-rich) are chosen for comparing the surface grand potential difference between the $CH_3NH_3Br$ and $PbBr_2$ surface terminations.

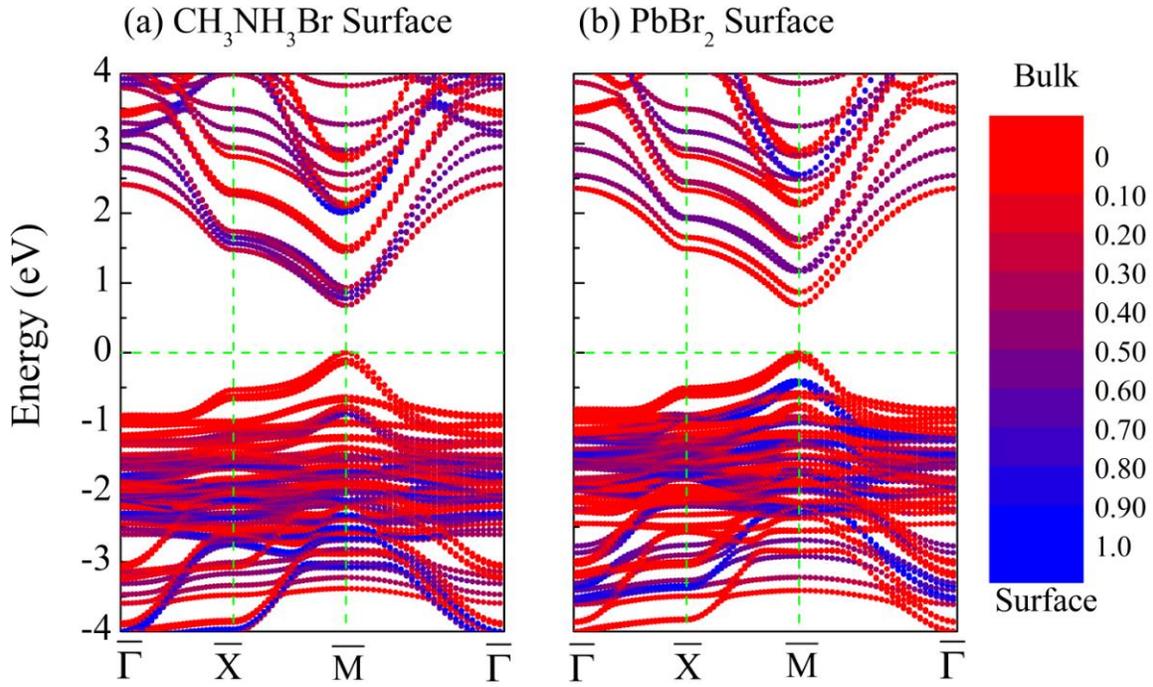

Fig. 4. The electronic band structure of the CH₃NH₃PbBr₃ slabs with the (a) CH₃NH₃Br and (b) PbBr₂ surface (001) terminations. The bands are shown along lines connecting high-symmetry points in the Brillouin zone with $\overline{\Gamma}$ (0,0,0), $\overline{X}$ (0,1/2,0), $\overline{M}$ (1/2,1/2,0). The bands are colored according to the relative contribution from bulk or surface, at a given energy. The surface (bulk) character is indicated by blue (red) color. The purple color implies that the surface and bulk bands strongly hybridized. The reference energy is placed at the CBM.

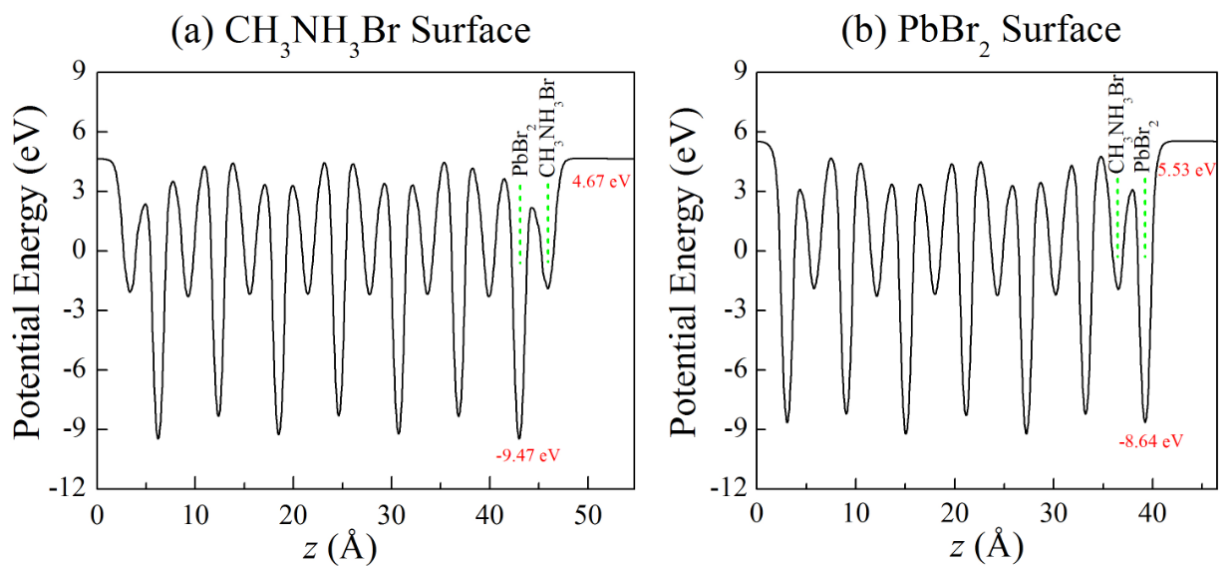

Fig. 5. The electrostatic potential energy profile across $CH_3NH_3PbBr_3$ (001) slabs, with the (a) $CH_3NH_3Br$ and (b) $PbBr_2$ surface terminations. The sharp dips in the potential energy correspond to $PbBr_2$ monolayers whereas shallower dips are due to $CH_3NH_3Br$ monolayers. The flat region corresponds to the vacuum energy. All the energies are referenced with respect to the VBM.

**Table 1.** The total energies $E_{tot}$ and heat of formations $\Delta H_{form}$ for various pertinent systems. All results are calculated with the Perdew-Burke-Ernzerhof (PBE) pseudopotential.[**Error! Bookmark not defined.**]

| System | $E_{tot}$ (eV/formula) | System | $E_{tot}$ (eV/formula) |
|---|---|---|---|
| $CH_3NH_3PbBr_3$ (cubic phase) | -52.792 | $Br_2$ (molecule) | -2.992 |
| $PbBr_2$ | -9.764 | C (graphite) | -9.221 |
| $CH_3NH_3Br$ (solid phase)[52] | -42.942 | $N_2$ (molecule) | -16.624 |
| Pb (metal, fcc lattice) | -3.711 | $H_2$ (molecule) | -6.760 |
| System | | $\Delta H_{form}$(eV) | |
| $CH_3NH_3PbBr_3$ | | -6.780 | |
| $CH_3NH_3Br$ | | -3.633 | |
| $PbBr_2$ | | -3.061 | |

**Table 2.** The calculated surface grand potential for the $CH_3NH_3Br$ surface termination and $PbBr_2$ surface termination at the representative points A, B and C in Fig. 3. Values of the surface grand potential are given in units of meV/$\text{Å}^2$ and J/m$^2$ in parentheses.

| | A (Pb-rich/Br-poor) | B (Moderate) | C (Pb-poor/Br-rich) |
|---|---|---|---|
| $CH_3NH_3Br$ Termination | 7.5 (0.120) | 8.0 (0.129) | 7.5 (0.120) |
| $PbBr_2$ Termination | 12.4 (0.197) | 11.8 (0.189) | 12.4 (0.197) |

**Table 3.** The calculated energy gaps (in units of eV) for bulk and (001) surfaces of $CH_3NH_3PbBr_3$.

|  | $CH_3NH_3PbBr_3$ polar bulk | $CH_3NH_3PbBr_3$ non-polar bulk | $CH_3NH_3Br$ termination | $PbBr_2$ termination |
|---|---|---|---|---|
| Band gap | 0.89 | 0.62 | 0.67 | 0.65 |